\documentclass{article}

\usepackage{arxiv}

\usepackage[utf8]{inputenc} 
\usepackage[T1]{fontenc}    
\usepackage{hyperref}       
\usepackage{url}            
\usepackage{booktabs}       
\usepackage{amsmath}
\usepackage{amsfonts}       
\usepackage{nicefrac}       
\usepackage{microtype}      
\usepackage{graphicx}
\usepackage{natbib}
\usepackage{doi}

\title{DCI: An Accurate Quality Assessment Criteria for Protein Complex Structure Models}

\author{ \href{https://orcid.org/0000-0003-2469-8522}{\includegraphics[scale=0.06]{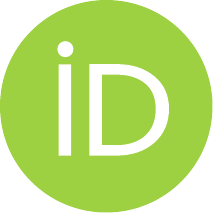}\hspace{1mm}Wenda Wang} \\
	Institute for Mathematical Sciences\\
	 Renmin University of China\\
	Beijing, China, 100872 \\
	\texttt{wangwenda87@ruc.edu.cn} \\
	\And
	{Jiaqi Zhai} \\
	Institute for Mathematical Sciences\\
	 Renmin University of China\\
	Beijing, China, 100872 \\
	\texttt{2022000744@ruc.edu.cn} \\
	 \AND
	 {He Huang} \\
	Institute for Mathematical Sciences\\
	 Renmin University of China\\
	Beijing, China, 100872 \\
	\texttt{hehuang@ruc.edu.cn} \\
        \And
	 *\hspace{1mm}{Xinqi Gong} \\
	Institute for Mathematical Sciences\\
	 Renmin University of China\\
	Beijing, China, 100872 \\
	\texttt{xinqigong@ruc.edu.cn} \\
}

\renewcommand{\shorttitle}{DCI: An Accurate Quality Assessment Criteria for Protein Complex Structure Models}

\hypersetup{
pdftitle={A template for the arxiv style},
pdfsubject={q-bio.NC, q-bio.QM},
pdfauthor={Wenda Wang, Jiaqi Zhai, HeHuang, Xinqi Gong},
pdfkeywords={Protein Complex Structures, Evaluation Metric,  Matrix Calculation},
}

\begin{document}
\maketitle

\begin{abstract}
\qquad The structure of proteins is the basis for studying protein function and drug design. The emergence of AlphaFold 2 has greatly promoted the prediction of protein 3D structures, and it is of great significance to give an overall and accurate evaluation of the predicted models, especially the complex models. Among the existing methods for evaluating multimer structures, DockQ is the most commonly used. However, as a more suitable metric for complex docking, DockQ cannot provide a unique and accurate evaluation in the non-docking situation. Therefore, it is necessary to propose an evaluation strategy that can directly evaluate the whole complex without limitation and achieve good results. In this work, we proposed DCI score, a new evaluation strategy for protein complex structure models, which only bases on distance map and CI (contact-interface) map, DCI focuses on the prediction accuracy of the contact interface based on the overall evaluation of complex structure, is not inferior to DockQ in the evaluation accuracy according to CAPRI classification, and is able to handle the non-docking situation better than DockQ. Besides, we calculated DCI score on CASP datasets and compared it with CASP official assessment, which obtained good results. In addition, we found that DCI can better evaluate the overall structure deviation caused by interface prediction errors in the case of multi-chains. Our DCI is available at \url{https://gitee.com/WendaWang/DCI-score.git}, and the online-server is available at \url{http://mialab.ruc.edu.cn/DCIServer/}.
\end{abstract}


\section{Introduction}
Quality assessment of predicted models is very important for protein structure prediction task. Giving an accurate score for the predicted protein structure will help a lot for the improvement of the prediction methods and the subsequent applications and studies.

In the evaluation metrics for an overall protein structure, the most traditional one is RMSD (Root Mean Square Deviation) \citep{1976A, Kabsch1978ADO}, which measures the deviation between the predicted model with the native structure by calculating the distance of equivalent atoms after structure superimposition. But, RMSD calculates the deviation of aligned atoms, if the predicted model is especially different from its native structure, the bad regions of non-aligned parts will be ignored, and RMSD measures the distance of all pairs of equivalent atoms equally, a small deviation will result in a high RMSD. TM-score (Template Model) \citep{TMscore} makes an improvement on the basis of RMSD. Instead of using simple root mean square deviation during superposition, TM-score uses an objective function that can eliminate the influence of protein sequence length on the score, and the distance of corresponding residues is scaled to obtain a continuous value between 0-1 as the final score. GDT \citep{PMID:12824330} was used to evaluate structures since CASP 4 \citep{CASP4}; its method is to calculate the maximum proportion of atoms whose distance after superposition is less than the threshold value in the total atoms according to the specified threshold value, and select the average of the proportion obtained by different thresholds as the final score. GDT can greatly reduce the influence of atoms that seriously deviate from the native structure on the evaluation results, and has a stricter requirement for the completeness of the prediction part. The algorithm to find the best Superimposition is complicated, and the structure can only be evaluated as a rigid body. More importantly, the evaluation strategy based on superimposition is strongly influenced by domain motions, changes in the relative orientation between domains may lead to large changes in the evaluation scores, while lDDT \citep{10.1093/bioinformatics/btt473}, as a local evaluation metric without superimposition, is more flexible and extensive in its application. lDDT considers pairs of atoms from different residues within a predefined distance threshold, calculates the distance difference between this pair of atoms in the native structure and the predicted model, and finally scores the average of the proportion of atom pairs with distance differences within 0.5\AA, 1\AA, 2\AA, 4\AA.

All of the above mentioned evaluation metrics are mainly used for protein monomer. But for complexes, the contact sites and interface between protein and protein are crucial for their structure and function \citep{IS}. In addition, with the emergence of AlphaFold 2 \citep{AF2, AF2-multimer}, single protein structures can be predicted almost accurate. Now when evaluating complexes, the quality of the contact interface is the most critical difference compared to the evaluation of monomers. In both CASP (Critical Assessment of Structure Prediction) and CAPRI (Critical Assessment of PRotein Interaction prediction) assessments, there is an evaluation strategy specifically for complexes. In recent years at the CASP, such as CASP 14 \citep{CASP14_1, CASP-CAPRI} and recent CASP 15, the official assessment criteria uses oligo-lDDT, TM-score, ICS (interface contact score) and IPS (interface patch score) to evaluate complex predicted results, where oligo-lDDT evaluating models locally, TM-score evaluating models globally, ICS and IPS evaluating interface areas specifically. The complex assessment strategy in CAPRI takes a different approach, which uses three metrics related to the contact interface \citep{CAPRI}: $F_{nat}$, lRMS and iRMS. $F_{nat}$ is the proportion of native contacts, lRMS is the backbone RMSD of the ligand common residues after the receptors are superimposed. iRMS is the backbone RMSD of the common interface residues. Based on these three metrics, the predicted model quality can be classified as high, medium, acceptable and incorrect.
DockQ \citep{DockQ} is a continuous model quality metric that combines $F_{nat}$, lRMS, and iRMS for protein-protein docking, currently the most versatile and recognized method for evaluating protein complexes, which has also been used in AlphaFold-multimer. DockQ can evaluate not only dimers but also multimer protein complexes. For evaluating multimers, DockQ divides all the chains of the multimer into two parts and then calculate the value just like that for a dimer. But because of that, as a suitable metric for docking, DockQ depends heavily on the selection of ligand and receptor and structure superimposition, so cannot provide an unique and accurate evaluation for a multimer complex without the knowledge of which chains are ligand or receptor. In conclusion, it is still necessary to develop an objective and good evaluation metric for evaluating entire protein complex structures.

\begin{figure}
\centering
\includegraphics[width=1\linewidth]{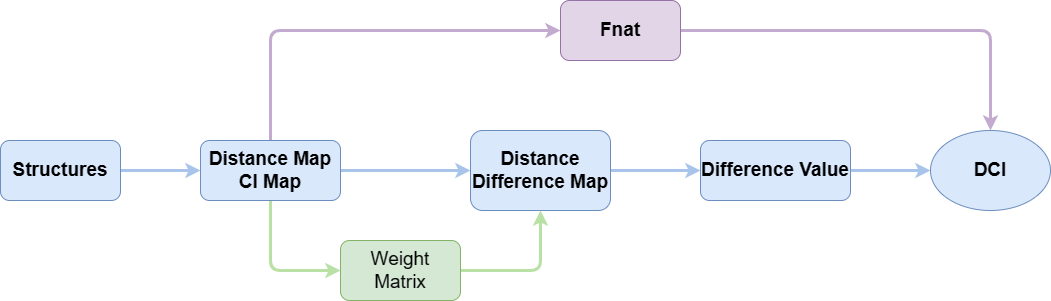}
\caption{\textbf{DCI pipeline.} Firstly, calculate distance map and CI map of two structures respectively. The distance maps are used to obtain distance difference map and the CI maps are used to calculate $F_{nat}$ and generate the weight matrix together with native structure. Distance difference map is multiplied by weight matrix and normalized to a difference value $V$. At last, $V$ is weighted sum with $F_{nat}$ to get DCI score.}
\label{fig:01}
\end{figure}

In this work, we developed DCI score, a new assessment metric for protein complex, which only based on distance map and CI (contact-interface) map, achieving good results both in comparison with DockQ and CASP assessment. The pipeline of DCI is shown in Fig.~\ref{fig:01}.

DCI highlights the accuracy of the contact interface based on the overall comparison of two complex structures. It takes a pair of structures to be compared as input, calculates the distance map and CI map of the two structures respectively. The CI map is used to calculate $F_{nat}$, to characterize intra-chain contacts and inter-chain contacts in the complex separately, we extend $F_{nat}$ to intra-$F_{nat}$ and inter-$F_{nat}$. The next step is comparing the distance maps to obtain distance difference map. Then we measure the importance of residues according to whether a pair of residues contact and which chain they are located respectively. We give a higher weight to the important residues pair and a lower weight to the unimportant residues pair to form a weight matrix which used to act on the distance difference map to get the difference value $V$. $V$ represents the overall difference between the predicted structure and the native structure, $F_{nat}$ represents the accuracy of contact prediction. So $V$ is weighted and summed with $F_{nat}$ to obtain the final DCI score.

\section{Materials and Methods}

\subsection{DCI maps}
When the three-dimensional structure of a protein complex is obtained, the distance map and the contact-interface map can be naturally calculated, in which the elements in the distance map represent the C$\alpha$-atom distance between any two residues in the complex. At the same time, contacts for intra-chain and interfaces for inter-chain can be obtained by applying different distance thresholds to the residue pairs located between or within a chain, and we made the inter-chain contact threshold slightly larger than the intra-chain contact threshold to obtain more interface contacts. In this work, the inter-chain contact threshold is set to 8\AA, and the intra-chain contact threshold is set to 6\AA, the resulting contact-interface map is a binary matrix of 1 if there is a contact and 0 if there is no contact.

\subsubsection{Fraction of Native Contact in Predicted Interface}
$F_{nat}$ in CAPRI\citep{CAPRI} indicates the fraction of native contacts preserved in the contacts of the predicted model. Because of the great improvement in the prediction accuracy of monomer, the prediction accuracy of intra-chain contacts in the complex has reached a high level, and inter-chain contacts are currently the main factor affecting the overall prediction accuracy of the structure. In order to characterize intra-chain contacts and inter-chain contacts in the complex separately, we extend $F_{nat}$ to intra-$F_{nat}$ and inter-$F_{nat}$, where inter-$F_{nat}$ represents the fraction of native inter-chain contacts appeared in the predicted model, and intra-$F_{nat}$ represents the fraction of native intra-chain contacts appeared in the predicted model.
\subsubsection{Distance Difference Map}
Obviously, two similar structures tend to have similar distance maps, so the starting point is to compare the difference of distance maps between structures to obtain their similarity. As shown in Equation.\ref{eq:01}, distance difference map $\widetilde{D}$ represents the absolute value of the difference between the distance map of ground truth $D_g$ and predicted model $D_p$. The element size in $\widetilde{D}$ represents the similarity of the position of corresponding residues pair in the structures, and a smaller value means that the two structures are more similar at that residues pair. Since distance matrix can be calculated by single structure, so the superimposition operation can be skipped after getting common residues of a pair of structures.
\begin{equation}
\widetilde{D}=|D_g-D_p|\label{eq:01}\vspace*{-5pt}
\end{equation}
\subsection{Mask and Weight Matrix}
\begin{figure}
  \centering
   \includegraphics[width=0.8\linewidth]{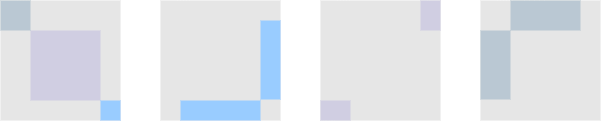}
   \caption{\textbf{Schematic diagram of masks. } The picture shows the masks of a trimer. The first mask on the left indicates whether the residues pairs are in the same chain. The colored part are intra-chain and the gray part are inter-chain. The three masks on the right indicate which two chains the residue pairs are located in when they are in different chains. For example, the dark blue part of the first mask on the right represents that the residue pairs come from the first and second chains respectively. For all masks, assign 1 to the colored parts and 0 to the gray parts.}
   \label{fig:02}
\end{figure}

It is not reasonable to consider the difference of each pair of residues with the same importance, so consider a weight matrix acting on the distance difference matrix. The value of the element in the weight matrix represents the importance of the distance difference of the corresponding position of the residues pair, which is determined by i) whether a pair of residues is in contact, ii) whether it is from intra-chain or inter-chain, and iii) from which chain respectively if from inter-chain. The objective of our consideration is to focus more on the inter-chain, contact residues. And if the residues pair are located in two different chains, the closer contact between the two chains have, the more emphasis on the evaluation of the residues pair from these two chains. To represent the location of the three types of residue pairs above, we define the corresponding mask matrices: The inter-chain/intra-chain mask matrix $M_I$ is a binary matrix where the corresponding position of the intra-chain residues pair is 1 and the corresponding position of the inter-chain residues pair is 0 (as Fig.~\ref{fig:02} left 1); For a chains set $\mathbb{C}$ of a structure, $M_{jk}$ is another binary matrix, the corresponding position of the residues pair from chain $j$ and chain $k$ is 1 ($j, k\in\mathbb{C}$), otherwise the corresponding position is 0 (as Fig.~\ref{fig:02} right 1-3); And the mask matrix $M_C$ which represents whether a residues pair is in contact, is exactly the native CI map $CI_g$.

The weight matrix $W$ measuring the importance of residues pair is combined by mask matrices mentioned above. The final weight matrix is composed of two parts: first we use $M_I$ to give a higher weight $\omega_{ie}$  for inter-chain residues pair and a lower weight  $\omega_{ia}$ for intra-chain residues pair. $W_I$ that give different weights to intra-chain and inter-chain residues pairs is defined as Equation.\ref{eq:02}. And we considered weighted average for intra-chain weight and inter-chain weight as $\omega_{ie} + \omega_{ia} = 1$.
\begin{equation}
    W_I=\omega_{ia}M_I+\omega_{ie}(1-M_I)\label{eq:02}
\end{equation}
And use $M_C$ to enlarge the weight of contact residues pair and reduce the weight of non-contact residues pair by two coefficients $\omega_c$ and $\omega_{nc}$ correspondingly. $W_C$ that give different weights to contact and non-contact residues pairs is defined as Equation.\ref{eq:03}. 
\begin{equation}
    W_C=\omega_c M_C+\omega_{nc}(1-M_C)\label{eq:03}
\end{equation}
second we use $M_{jk}$ to give different weight for residues pair from different chains, that the weight $\omega_{jk}$ will be higher if two chains have more contacts, to achieve this, we define a series of interface features: for an interface formed by chain j and chain k, $L_j. L_k$represents the respective length of the two chains, $I_{kl}$ is the number of interface residue pairs, While $C_j$ and $C_k$ represent the number of residues in the interface of the two chains, and $C_{jk}$ represents the total number of residues in the interface, i.e. $C_{jk}=C_j+C_k$. We define $\omega_{jk}=\frac{I_{jk}^2+C_{jk}^2}{L_j \cdot L_k}$ as the weight that measures the importance of the contact between the two chains. $w_D$ that assign different weights to residues from different chains is shown as Equation.\ref{eq:04}.
\begin{equation}
    W_D=\sum\limits_{j, k\in \mathbb{C}}\omega_{jk}M_{jk}\label{eq:04}
\end{equation}
The final weight matrix $W$ is obtained based on the expansion or reduction of $W_I$ and $W_C$, plus the additional weight of $W_D$ as Equation.\ref{eq:05}.
\begin{equation}
    W=W_IW_C+W_D\label{eq:05}
\end{equation}
After obtaining distance difference matrix and weight matrix, the Hadamard product of the two matrices is mapped to the difference value $V$ on the 0-1 continuous interval as shown in Equation.\ref{eq:06}.

{\setlength\abovedisplayskip{3pt}
\setlength\belowdisplayskip{13pt}
\begin{equation}
V=\frac{1}{1+\overline{W\odot\widetilde{D}}}\label{eq:06}
\end{equation}}

So far we have gotten V represents the overall difference between the predicted structure and the native structure and intra-$F_{nat}$, inter-$F_{nat}$ represents the accuracy of intra-chain and inter-chain contact prediction. Finally DCI score is obtained by weighted average of intra-$F_{nat}$, inter-$F_{nat}$ and difference value $V$ as shown in Equation.\ref{eq:07}, where $\omega_1 + \omega_2 + \omega_3 = 1$.
\begin{equation}
DCI=\omega_1\cdot Intra\operatorname{-}F_{nat}+\omega_2\cdot Inter\operatorname{-}F_{nat}+\omega_3\cdot V\label{eq:07}
\end{equation}

\begin{figure}
  \centering
   \includegraphics[width=0.8\linewidth]{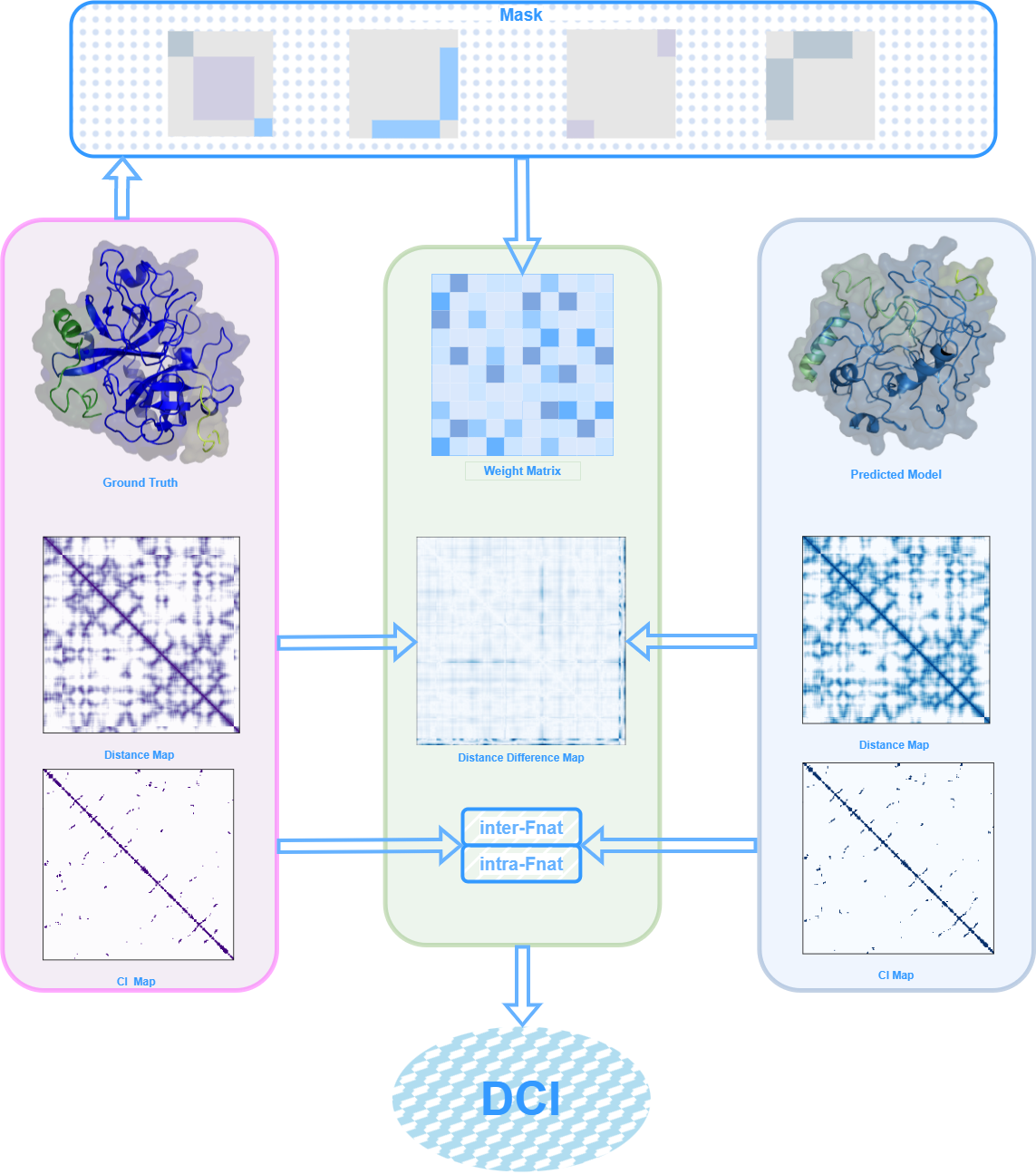}
   \caption{\textbf{algorithm of DCI. } DCI inputs a set of structures for comparison and calculates the distance map and CI map for each structure (purple represents the true structure and blue represents the predicted structure, dark regions in the maps represent the close or contact of residue pairs). Distance difference map is calculated using the distance maps of two structures, and $F_{nat}$ is calculated using true CI map, several masks of the position of residues pair was generated according to the native structure, which is combined into the weight matrix and operate Hadamard product with the distance difference map, DCI score is obtained by weighted summation of this value with $F_{nat}$ finally.}
   \label{fig:03}
\end{figure}

The diagram containing the above algorithm flow is shown in Fig.~\ref{fig:03}.

\subsection{Penalty}
In order to make the distance matrix of the native structure and the predicted structure can be subtracted, it is required to keep the dimension of the two matrices consistent, so it is necessary to intercept the residues common to the two structures. If the residues of the native structure are cropped due to the loss of residues in the predicted structure, a penalty coefficient will be added to the DCI calculation. The penalty coefficient is the ratio of the cropped length to the original length in native structure. In practice, whether calculating DCI with penalty is optional.
\subsection{Datasets}
Here we use five protein complex datasets for parameter optimization and results analysis. We briefly described each dataset and labels for evaluation and comparison. More detailed information about the datasets is presented in the Appendix.
\subsubsection{Overview of Five Datasets}
\begin{enumerate}
\item \textbf{CAPRI set}\citep{DockQ} contains models submitted to CAPRI between 2005–2014 with their respective DockQ scores and CAPRI classifications. Since the prediction models of T35 and T36 contains only one chains in the pdb files, we removed these two targets, then it has 13 targets and 18205 predicted models totally. According to CAPRI classification, there are 16042 incorrect, 860 acceptable, 817 medium, 486 high models. (\url{http://cb.iri.univ-lille1.fr/Users/lensink/Score\_set/}) 
\item \textbf{Heterodimer-AF2 (HAF2) set} was generated by collecting the structures of heterodimers from the Protein Data Bank and predicting different models for each heterodimer (\url{https://zenodo.org/record/6569837}). Overall, the HAF2 dataset has 13 targets with 1370 models, 457 incorrect, 84 acceptable, 488 medium and 341 high.
\item \textbf{Multimer-AF2(MAF2) set} used AlphaFold 2 and AlphaFold-Multimer to generate protein complex structures for protein targets derived from the EVCoupling \citep{EVcoupling} and DeepHomo \citep{10.1093/bib/bbab038} datasets.(\url{https://zenodo.org/record/6570843}) There are 2503 targets with 4985 models have correct classification and DockQ value.
\item \textbf{CASP 14 set} includes multimers in the CASP 14 whose native structures can be found, there are 17 targets and 1914 models. (\url{https://www.predictioncenter.org/casp14})
\item \textbf{CASP 15 set} includes multimers in the CASP 15 whose native structures can be found and official assessments are shown on CASP website, there are 21 targets and 5304 models. (\url{https://www.predictioncenter.org/casp15/})
\end{enumerate}
\subsubsection{Reference Label}
In order to complete the parameter optimization and result testing with the above datasets, a reference label needs to be provided for each model. Labels in the datasets can be divided into two categories: classification label and regression label. On the one hand, CAPRI dataset and HAF2 dataset contain both DockQ result and CAPRI classification, MAF2 dataset only has DockQ result but can calculate CAPRI classification for dimer models (multimers more than two chains need assign ligand and receptor to get CAPRI classification), then the CAPRI classification in these three datasets can be regarded as reference label in the next optimization. On the other hand, CASP14 dataset has several evaluation metrics supplied by CASP, we select four main metrics that used in official assessment of recent CASP competition---oligo-lDDT, TM-score, ICS (F1 score) and IPS (Jaccard Coefficient) and calculate mean value of these four metrics as the reference label. According to the type of reference label, we can also divide datasets into classification set (CAPRI, HAF2, MAF2) and regression set (CASP14, CASP15), totally 2567 targets and 31778 models,  the usage and number of obtained datasets are shown in Tabel~\ref{Tab:01}.
\begin{table}
    \renewcommand\arraystretch{1.2}
    \centering
    \caption{\centering Description of dataset}
    \begin{tabular}{cccc}
    \hline
        \textbf{Type} & \textbf{Dataset} & \textbf{Target} & \textbf{Model} \\ \hline
        \textbf{} & CAPRI & 13 & 18205 \\ 
        \textbf{Classification} & HAF2 & 13 & 1370 \\ 
        \textbf{} & MAF2 & 2503 & 4985 \\ \hline
        \textbf{Regression} & CASP14 & 17 & 1914 \\ 
        \textbf{} & CASP15 & 21 & 5304 \\
        \textbf{Summary} & ~ & 2567 & 31778 \\ \hline
    \end{tabular}
    \label{Tab:01}
\end{table}

\subsection{Parameter optimization}
The process of DCI calculation mentioned in "Mask and Weight Matrix" requires 7 parameters including inter-chain weight $\omega_{ie}$, intra-chain weight $\omega_{ia}$ in Equation.\ref{eq:02}, contact weight $\omega_c$, non-contact weight $\omega_{nc}$ in Equation.\ref{eq:03}, and DCI weight coefficients $\omega_1, \omega_2, \omega_3$ in Equation.\ref{eq:07}. And we have two constraints in Equation.\ref{eq:08}:
{\setlength\abovedisplayskip{0pt}
\setlength\belowdisplayskip{15pt}
\begin{equation}
\omega_{ie}+\omega_{ia}=1\qquad \omega_1+\omega_2+\omega_3=1\label{eq:08}\vspace*{-9pt}
\end{equation}}
so that the number of weight coefficients can be reduced to 5: $\omega_{ie}, \omega_c, \omega_{nc}, \omega_2, \omega_3$. In addition to optimizing these weight coefficients, we optimized classification cutoffs to calculate CAPRI classification by DCI score as accurate as possible. Here we give a brief introduction about the process and result of parameter optimization, more details are mentioned in the Appendix.
\subsubsection{Weight coefficient}
In this work we used Bayesian optimization \citep{Frazier2018ATO} on five weight coefficients, which aims at finding an acceptable local optimal solution for an objective function $f(\pmb{\omega}|\pmb{x}, y)$, where $\pmb{\omega}=(\omega_{ie}, \omega_c, \omega_{nc}, \omega_2, \omega_3)$, $\pmb{x}$ and $y$ represent datasets and their labels. we use a few objective functions $f$ to find more possible weight coefficient combinations including overall $MCC$ (Matthews Correlation Coefficient), $w\_MCC$ (weighted MCC for four classifications) of CAPRI classification, $cor$ (Pearson correlation coefficient), $MSE$ (square loss between DCI score and CASP reference value). We created training set by classification set and mixed set (both classification and regression set).After obtaining good parameter combinations in training set, we calculated $ACC$ (accuracy) and $MCC$ in test set and selected the one with the best performance. The size ratio of the training set to the test set is 8:2. The specific selection of objective functions and training set, and the parameter combinations selected from the training set can be found in Appendix. Finally we determined all weight coefficients as $\omega_{ie}=0.8, \omega_c=1.4, \omega_{nc}=0.9, \omega_2=0.17, \omega_3=0.8$.

\subsubsection{Classification cutoffs}
We expect to obtain the correct CAPRI classification by DCI score. After calculating the DCI score with the optimized weight coefficients, three appropriate cutoffs $c_1, c_2, c_3$ are selected to classify the classification data in the test set, so that the classification is as accurate as possible. The minimum step size of grid search is set to 0.01 in the optimization method, and $c_1, c_2, c_3$ are searched in the range of $[0.25, 0.5], [0.45, 0.7], [0.6, 0.85]$ respectively. Finally we selected $c_1=0.48, c_2=0.6, c_3=0.82$ as classification cutoffs.

\section{Results}
We have obtained the complete calculation method of DCI score according to Equation.\ref{eq:01}-\ref{eq:07}, and got the optimized parameter values and cutoffs, here we will show the excellence of DCI by comparing it with other common metrics in complex structure assessment.
\subsection*{Classification results}
\begin{figure*}[!htb]
  \centering
   \includegraphics[width=0.7\linewidth]{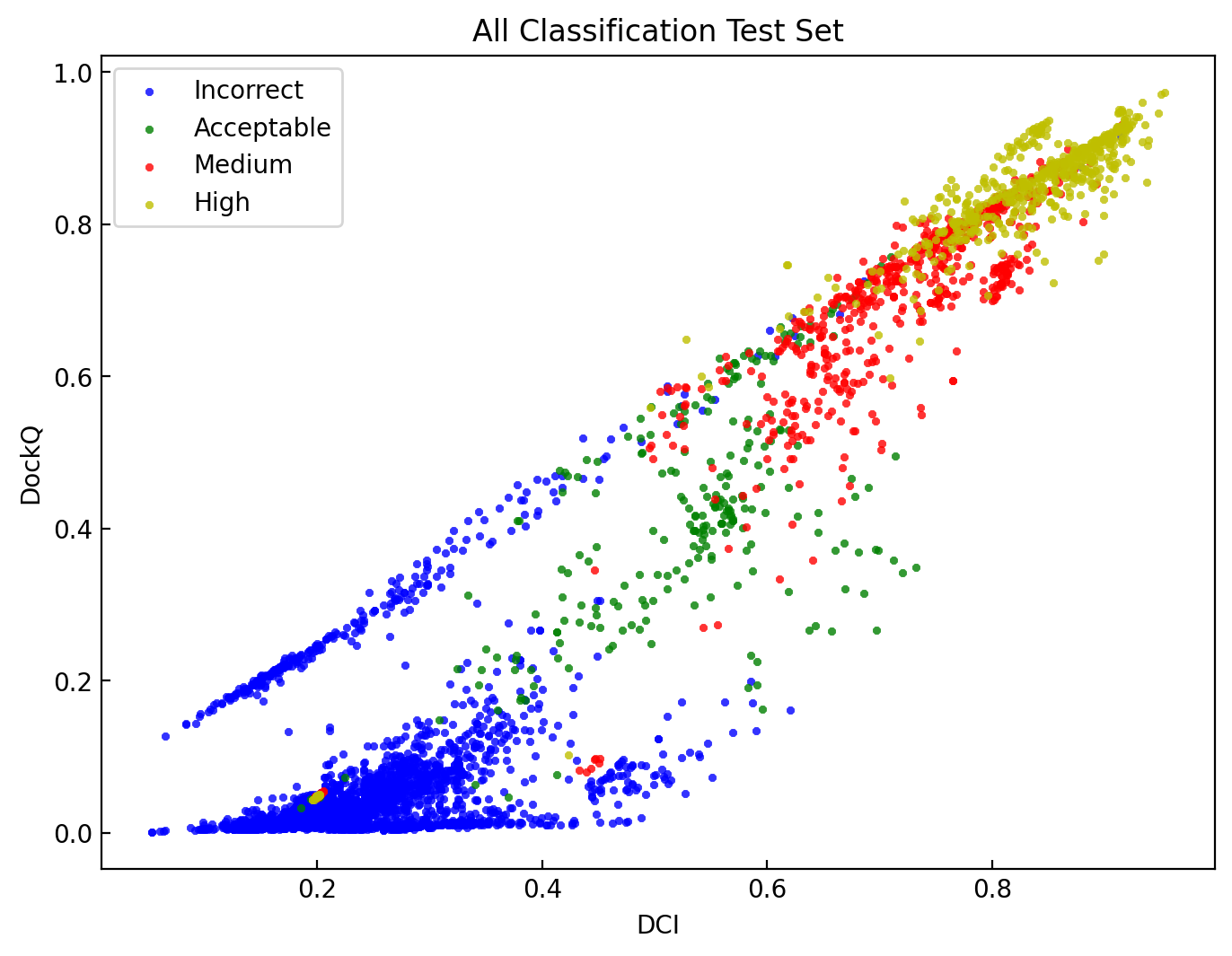}
   \caption{\textbf{Scatter plot of DCI versus DockQ on the test set.} Models are colored according to CAPRI classification as Incorrect (blue), Acceptable (green), Medium (red), High (yellow), and the Pearson correlation within the different quality classes is 0.46 (incorrect), 0.64 (acceptable), 0.9 (medium), 0.96 (high) respectively, and the overall Pearson correlation is 0.96.}
   \label{fig:04}
\end{figure*}

\begin{figure}[!htb]
  \centering
   \includegraphics[width=1\linewidth]{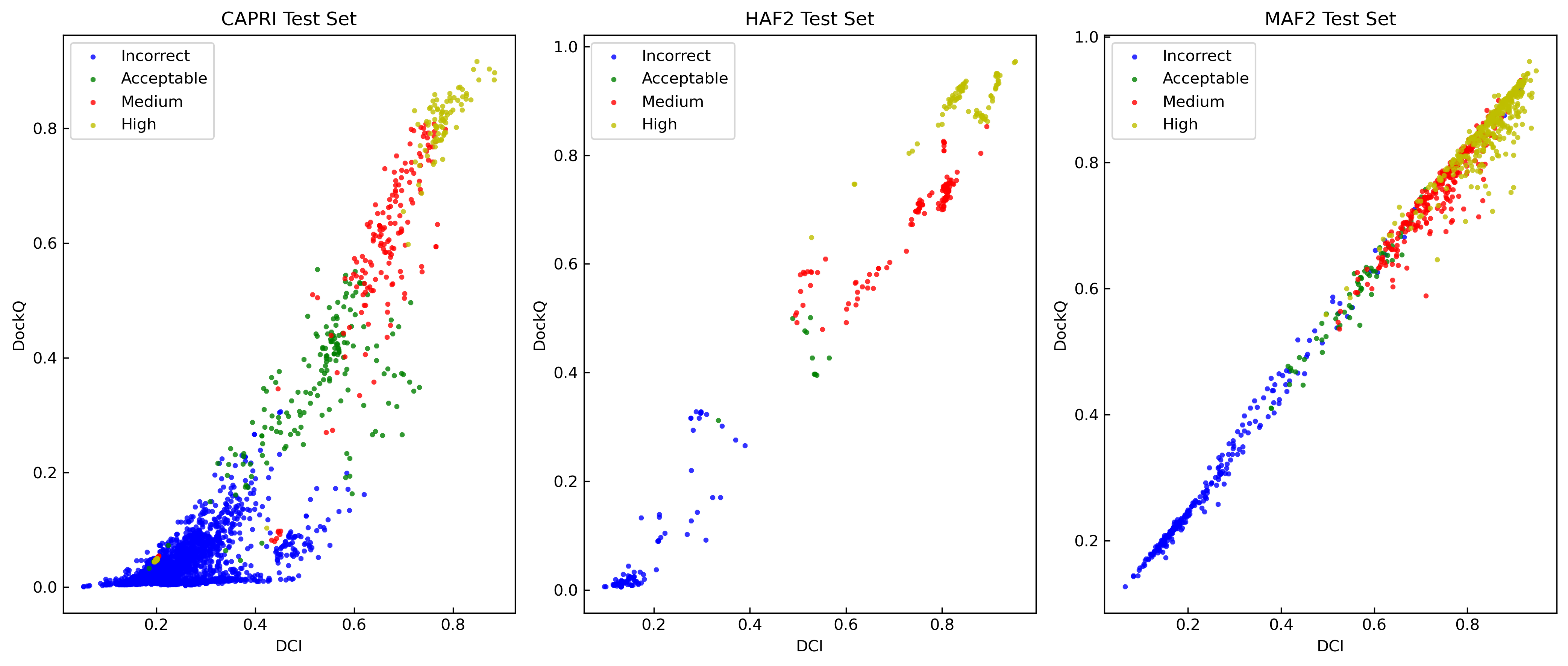}
   \caption{\textbf{Scatter plot of DCI vs. DockQ on each sub-test set.} Models are colored according to CAPRI classification as Incorrect (blue), Acceptable (green), Medium (red), High (yellow), the Pearson correlation between DCI and DockQ on each sub-test set is 0.91 (CAPRI), 0.98 (HAF2), 0.996 (MAF2) respectively.}
   \label{fig:05}
\end{figure}

This part of results is the comparison with DockQ. We evaluated the results not only on the complete test set, but also on the samples belonging to CAPRI, HAF2 and MAF2 in the test set which we called sub-test sets. First in terms of correlation, the overall Pearson correlation between DCI and DockQ is as high as 0.96, which indicates DCI score and DockQ, the state-of-the-art metric in complex assessment, have a strong connection. Fig.~\ref{fig:04} shows a scatter plot of DCI versus DockQ on the test set, 
and Fig.~\ref{fig:05} shows DCI versus DockQ on each sub-test set, the Pearson correlation between DCI and DockQ on CAPRI, HAF2, MAF2 sub-test set is 0.91, 0.98, 0.996 respectively.

In terms of comparison of CAPRI classification results, DCI score and DockQ results are equivalent. The correlation coefficients between DCI and CAPRI true labels, DockQ and CAPRI true labels are equal, both are 0.78. Moreover, for the classification results of the test set samples, DCI was better than DockQ in the overall results, both ACC (accuracy) and MCC (Matthews correlation coefficient), and the accuracy of the four categories was also equivalent to DockQ. Tabel~\ref{Tab:02} shows the specific classification accuracy results.
\begin{table}
    \centering
    \caption{\centering ACC and MCC comparation on test set}
    \begin{tabular}{@{}llll@{}}
    \hline
        \textbf{Target} & \textbf{Range} & \textbf{DCI} & \textbf{DockQ} \\ \hline
                        & overall        & 0.903        & 0.898          \\
                        & Incorrect      & 0.97         & 0.955          \\
        \textbf{ACC}    & Acceptable     & 0.96         & 0.949          \\
                        & Medium         & 0.929        & 0.935          \\
                        & High           & 0.948        & 0.956          \\ \hline
                        & overall        & 0.786        & 0.783          \\
                        & Incorrect      & 0.925        & 0.893          \\
        \textbf{MCC}    & Acceptable     & 0.557        & 0.518          \\
                        & Medium         & 0.692        & 0.701          \\
                        & High           & 0.718        & 0.783          \\ \hline
    \end{tabular}
    \label{Tab:02}
\end{table}

DCI showed comparable classification performance to DockQ on all three sub-test sets, we chose the MAF2 sub-test set with the best results for presentation as Tabel~\ref{Tab:03}. The other two sub-test sets' results are presented in the Appendix.

\begin{table}
\centering
\caption{\centering ACC and MCC comparation on MAF2 sub-test set}
\begin{tabular}{@{}lllll@{}}
\toprule
\textbf{Dataset} & \textbf{Target} & \textbf{Range} & \textbf{DCI} & \textbf{DockQ} \\ \midrule
\textbf{}        & \textbf{}       & overall        & 0.798        & 0.655          \\
\textbf{}        & \textbf{}       & Incorrect      & 0.972        & 0.861          \\
\textbf{}        & \textbf{ACC}    & Acceptable     & 0.943        & 0.826          \\
\textbf{}        & \textbf{}       & Medium         & 0.823        & 0.775          \\
\textbf{MAF2}    & \textbf{}       & High           & 0.857        & 0.847          \\ \hline
\textbf{}        & \textbf{}       & overall        & 0.713        & 0.52           \\
\textbf{}        & \textbf{}       & Incorrect      & 0.926        & 0.628          \\
\textbf{}        & \textbf{MCC}    & Acceptable     & 0.506        & 0.009          \\
                 &                 & Medium         & 0.604        & 0.489          \\
                 &                 & High           & 0.692        & 0.681          \\ \hline
\end{tabular}
\label{Tab:03}
\end{table}

\begin{figure}
  \centering
   \includegraphics[width=0.8\linewidth]{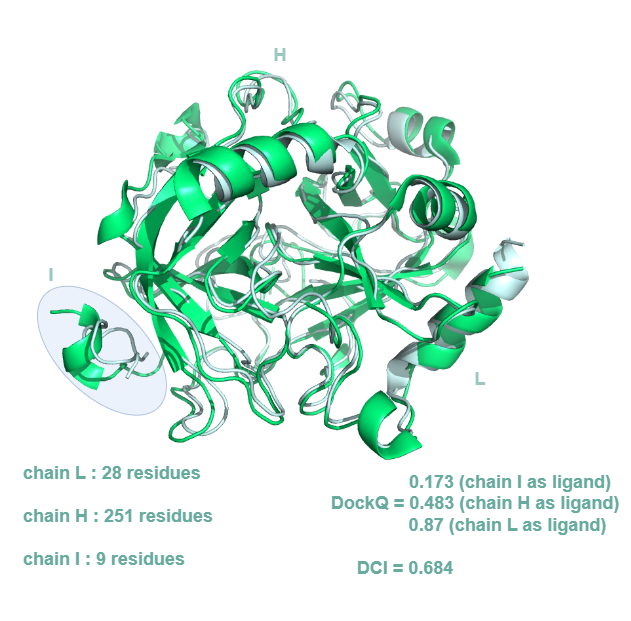}
   \caption{\textbf{Case study of complex 7AC9.} This complex has 3 chains but we do not know which part is the ligand and which part is the receptor, if choosing the shortest chain (I) as the ligand, the DockQ value is 0.173, and if calculating mean DockQ value when three chains are chosen as ligand in turn (0.173 for chain I, 0.483 for chain H and 0.87 for chain L), the value is 0.509, however, the predicted structure only has obvious errors on chain I, and the rest positions are basically consistent with the native structure. Compared with the two values calculated by DockQ, DCI = 0.684 can more correctly reflect the prediction accuracy of this complex.}
   \label{fig:06}
\end{figure}
\subsection{Advantages in multimer assessment}
\begin{figure}
  \centering
   \includegraphics[width=1.0\linewidth]{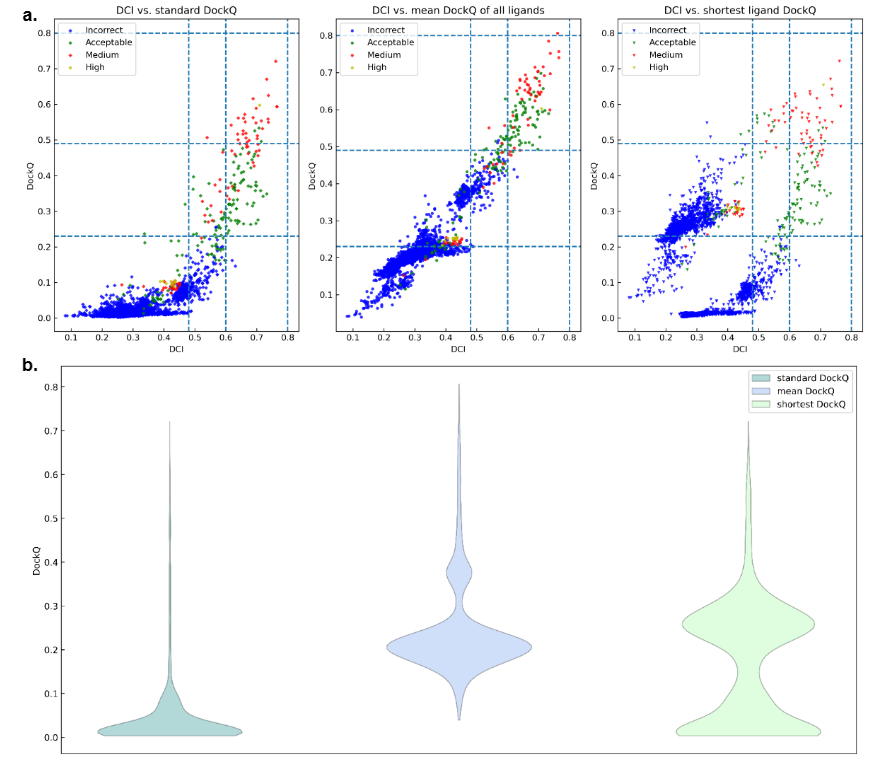}
   \caption{\textbf{DCI vs.multi-method DockQ on two CAPRI trimer targets.}  \textbf{a.} shows scatter plots of Target37 and Target40 DCI scores versus DockQ scores. The left subgraph is the standard case that calculate DockQ based on the known ligand. The latter two subgraphs show that when we assume that the ligand partition of the two targets is unknown, calculate DockQ with each chain as ligand and then average them (subgraph2), or calculate DockQ with the shortest chain in the structure as ligand (subgraph3). It can be seen that compared with the standard DockQ in subgraph1, the classification accuracy is significantly decreased in other two cases (ideally, the samples of the four classes should be located in the four diagonal grids). \textbf{b.} shows DockQ distribution differences calculated in these three cases. It is obviously that when ligand is unknown, the results of DockQ are unstable. So it can be seen that DockQ is not able to accurately evaluate the complex in the non-docking situation.}
   \label{fig:07}
\end{figure}

Here we show the advantage of DCI score in evaluating multimers compared to the current state-of-art metric DockQ, using a typical case (pdbID: 7AC9) as Fig.~\ref{fig:06}, an ordinary trimer with unknown ligand (required to calculate DockQ). If one of the chains is chosen as the ligand, It can be seen that the DockQ values of the three chains as ligand are very different. If we follow the current default principle of selecting ligand: choosing the shortest chain as the ligand, Then the ligand of this structure will be chain I which has only 9 residues, but obviously the predicted model is pretty similar to ground truth, and there is only a distinct deviation on the I chain, so it can be seen that DockQ, in the case of this unknown ligand, The evaluation of protein complexes is not sufficiently accurate. We also experimented with two trimer targets in the CAPRI dataset, with a total of 2950 samples, The results in Fig.~\ref{fig:07} show that when the ligand is unknown, the relevant calculation of DockQ cannot guarantee the correct evaluation. Whether it is to calculate DockQ separately with each chain as ligand and then take the average, or to choose the shortest chain as ligand by default, Both results in a significant decrease in the accuracy of structure evaluation (the horizontal and vertical dashed lines in the figure represent the classification thresholds of DockQ and DCI respectively), which also reflects the advantage of DCI score in directly evaluating multimer regardless of ligand selection. Through the above two sets of experiments, we clarify the instability of DockQ in calculating non-docking situation, and also show the advantage of DCI score in this ubiquitous situation.
\subsection{CASP results}
\begin{figure}
  \centering
   \includegraphics[width=0.8\linewidth]{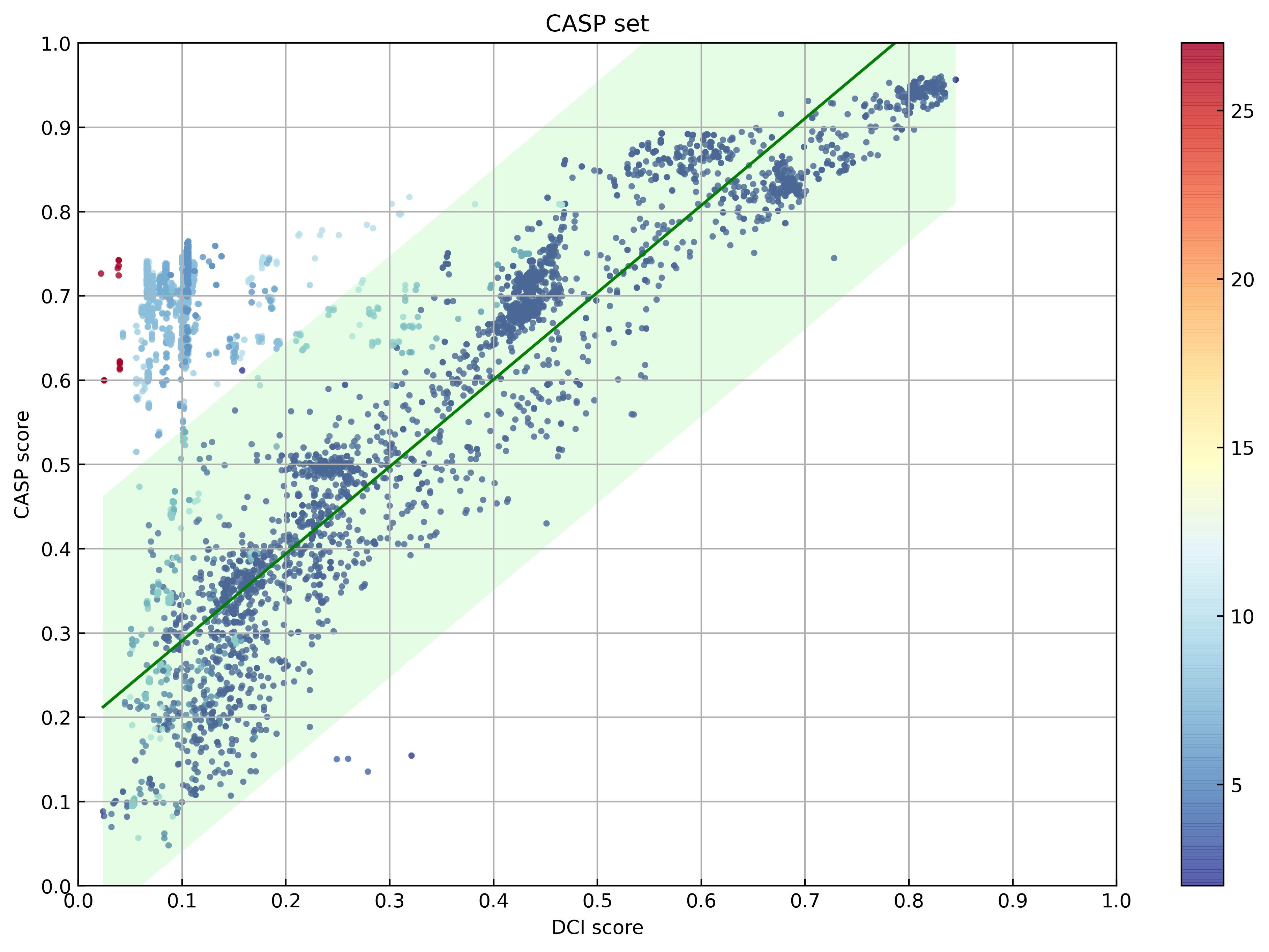}
   \caption{\textbf{Scatter plot of DCI versus CASP score on CASP set.} This figure contains 381 samples in CASP14 test set and 5314 samples in CASP15 set, color from cold to warm tones indicates the number of chains from small to large. The Pearson correlation coefficient between DCI score and CASP score is 0.522. The green line represents the fitting result of DCI score and CASP score when the number of chains is low (\textless 6), and the corresponding light green areas are the confidence interval of $\pm0.25$ given by us.}
   \label{fig:08}
\end{figure}

In this section, we evaluated the results of the CASP14 test set and the CASP15 set (hereafter referred to as a CASP set), then analyzed the batch interface errors appeared in CASP assessment and the performance of DCI score in CASP targets. Fig.~\ref{fig:08} shows the scatter plot of the DCI score and the CASP score :(oligo-lDDT+TM+ICS+IPS)/4.

 It can be found in Fig.~\ref{fig:08} that when the number of chains is small, DCI correlates 
strongly with CASP score, but when the number of chains increases, the samples will deviate from the confidence interval. The Pearson correlation coefficient of DCI and CASP score on the CASP set is 0.522, after removing samples with 6 or more chains from CASP set, the Pearson correlation coefficient increased to 0.942. We fitted the samples after removing models that more than 6 chains and generate $\pm0.25$ confidence intervals as overall trend. For outliers deviating from the overall trend, we divided them into the following two kinds.
 
 First kind of outliers with high CASP score and low DCI score at the left upper side of Fig.~\ref{fig:08}, whose chain numbers are large. Low DCI score versus high CASP score means DCI is more strict than CASP evaluation for large complexes. To find out the reasons, we first analyzed the four metrics in CASP score.
 
\begin{figure}
  \centering
   \includegraphics[width=1\linewidth]{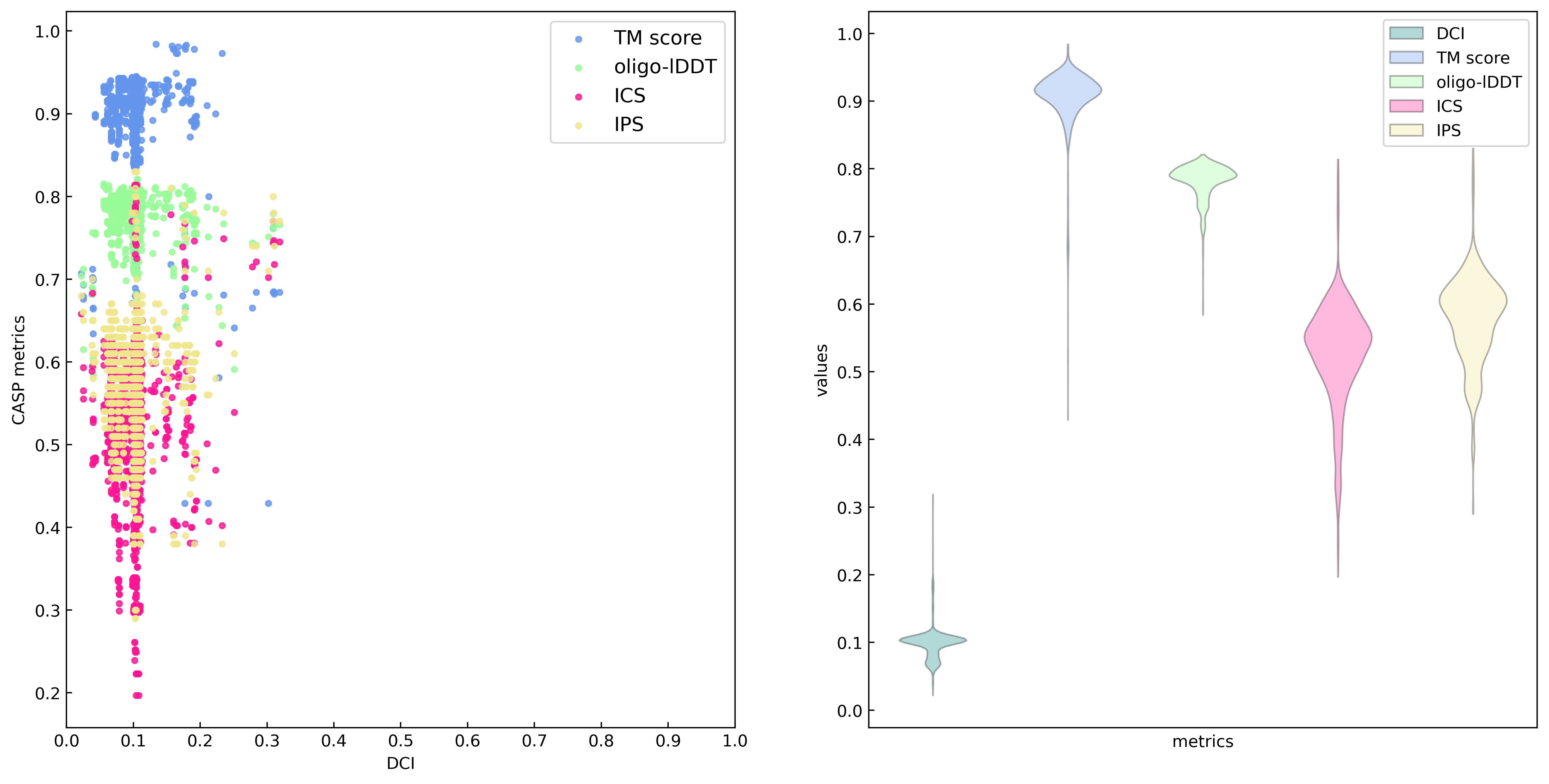}
   \caption{\textbf{Relationship between DCI score and four metrics in CASP score.} This figure shows the outliers above the overall trend in Fig.~\ref{fig:08}. The vast majority are complexes with more than 6 chains. The left picture shows the scatter plots of DCI score an four metrics (TM-score, lDDT, ICS, IPS) in CASP score respectively. It can be found that when the metrics measuring the interface (ICS, IPS) increases, the metrics measuring the overall structure (TM-score, lDDT) is almost horizontal and always at high values. And the right picture shows the difference in distribution between DCI score and the four metrics, when the number of chains is large, the strictness of DCI score compared with CASP score is mainly reflected in the difference between DCI score and the overall metrics TM-score and lDDT.}
   \label{fig:09}
\end{figure}

We analyzed the relationship between the four metrics in CASP score and DCI score as shown in Fig.~\ref{fig:09} focusing on the first kind of outliers. It can be found that when the metrics measuring the interface (ICS, IPS) increases, the metrics measuring the overall structure (TM-score, lDDT) is almost horizontal and always at high values. We consider that as the metrics mainly designed to evaluate single protein, TM-score and lDDT regard multimers as monomers when evaluating, which will cause inaccuracy in the overall structure evaluation of multimers. For example, lDDT is not sensitive to the relative orientation of domains, which is no longer an advantage when evaluating multimers because the relative orientation of chains is crucial to the multimers structures.

  When evaluating complexes using the metrics used to evaluate monomers, it is difficult to capture the overall structural deviation due to errors of interface prediction, but interface predicted accuracy is the most important for complex prediction compared to single chain prediction. Thanks to AlphaFold 2, the 3D structure of a single chain in a complex has already gotten high quality \citep{casp-af2}, but the relative position and orientation of the chains in a complex are mainly determined by the interface predicted accuracy. Even two correct single-chain structures, if connected by a wrong interface, the obtained overall structure also has a large deviation from the native structure. Moreover, the more chains there are, the more interface errors are likely to occur, then the more chains are likely to deviate from the correct position. We call this kind of problem : batch interface errors.

 Unlike CASP score, DCI can better reflect the overall structural deviation caused by batch interface errors. Because the mechanism of DCI is emphasizing the prediction accuracy of interface on the basis of evaluating overall structural similarity. The relative position deviation caused by interface error between two chains can be well reflected by the increase of corresponding elements in the distance difference map.

\begin{figure}
  \centering
   \includegraphics[width=1\linewidth]{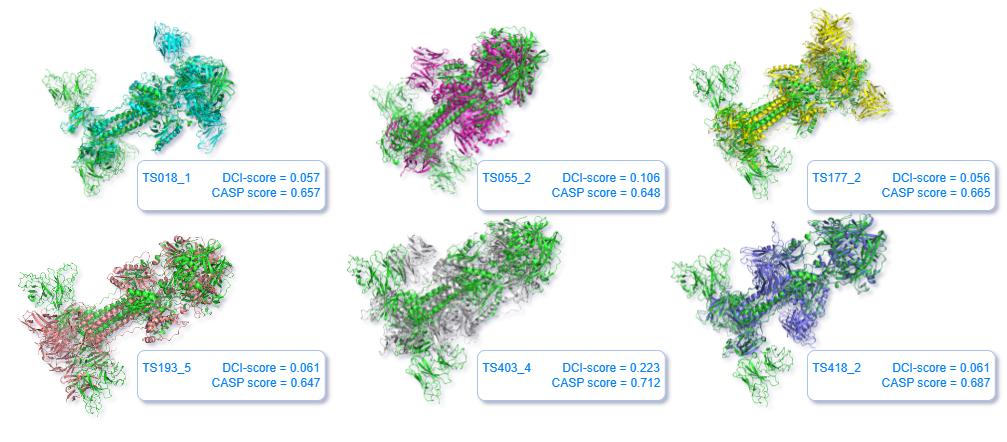}
   \caption{\textbf{Case study of target H1036.} This target has nine chains, the above models are the top six CASP scores among the predicted models of H1036. But after aligning these models with the native structure, it can be seen that the overall similarity of each prediction model is not high, And the main reason is that the predicted error of the interface formed by any neighboring chains would add up to a large deviation in the overall structure of the model. For example, some chains of TS018\_1 and TS177\_2 are completely opposite to the correct position, while some chains of TS055\_2, TS418\_2 and TS193\_5 are significantly deviated from the correct position. However, chains like TS403\_4  have relatively fewer deviation errors, so its DCI score (0.223) is higher than other models, which the selection of best model is consistent with the result of CASP assessment.}
   \label{fig:10}
\end{figure}

\begin{figure}
  \centering
   \includegraphics[width=0.8\linewidth]{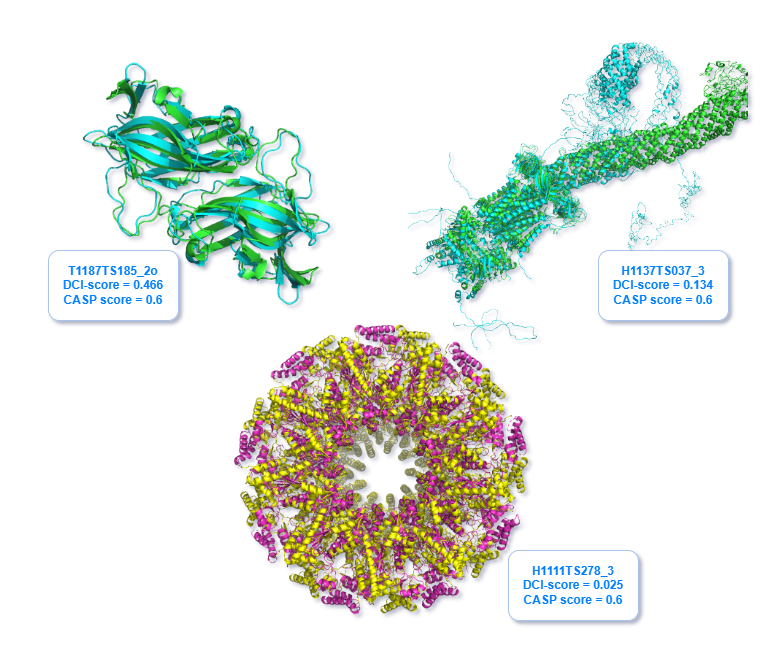}
   \caption{\textbf{3 Cases study of CASP15.} The above three pairs of structures were from T1187o (2 chains), H1137 (10 chains), H1111 (27 chains) and one predicted model of each, and the CASP score is 0.6, but as the number of chains increased, For example, the predicted structure of T1187o only shows many deviations in the loop region, while 6 chains of H1137 show serious deviations in direction. After alignment of the predicted model onto the native structure of H1111, the native structure and each chain of the predicted model show a complete "interleaved" state. Almost no single pair of chains can be broadly aligned in the overall structure. The "interleaved" details of H1111 is shown in Appendix.}
   \label{fig:11}
\end{figure}

 There are two sets of cases of CASP results that show the accuracy of DCI in evaluating multimers compared to CASP score. Fig.~\ref{fig:10} shows the top 6 CASP score cases of target H1036 (9 chains) in CASP14 test set, but the DCI scores are very low. 
Also as shown in Fig.~\ref{fig:11}, three predicted models have similar CASP scores around 0.6, but as the number of chains increases from 2 to 10 and then to 27, contact errors between interfaces lead to more and more differences in the relative orientation of chains, which leads to briefly decreases in DCI score.

According to the above analysis, compared with CASP assessment, DCI can better reflect the case of batch interface errors. And because the larger the number of chains, the more likely and number of batch interface errors will occur, so the DCI score will be more strict than the CASP score, that is, the more likely to have lower DCI score. 

It is worth noting that more attention is paid to the relative orientation between the chains, which is consistent with the situation that in the current structure prediction task, the single-chain structure prediction is very accurate, and the key difficulty is the combination of multi-chains.

\begin{figure}
  \centering
   \includegraphics[width=1\linewidth]{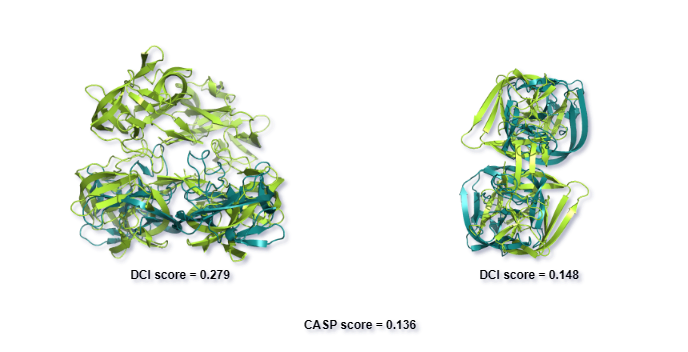}
   \caption{\textbf{Missing chain case in CASP set.} This case shows T1034o in CASP14, which represents the second kind of outlier cases. Its native structure has 4 chains, while the predicted model has only 2 chains. Therefore, in the process of structure length alignment, the first two chains of the native structure are clipped by default. However, the predicted model actually corresponds to the second and third chains in the native structure. As a result, DCI score was only 0.279. When the number and order of chains correspond correctly, the DCI score (0.148) returns to normal trend.}
   \label{fig:12}
\end{figure}

Second kind of outliers with low CASP scores and high DCI score, this case reflects a common problem in the use of most evaluation metrics, that is, the prediction models only have partial chains of the native structure. As shown in Fig.~\ref{fig:12}, This target is a proxy below the overall trend in Fig.~\ref{fig:08}, which DCI score is 0.279, higher than the CASP score of 0.136. This is the case because the native structure has 4 chains, but the predicted model contains only 2 of them. So when the length of these two structures is aligned, the first two chains of the native structure are selected by default. To solve this problem, the number and order of chains need to be determined at the time of input to facilitate structure alignment. If the number of chains does not correspond, it needs to be corrected manually. If the chain order does not correspond, our program provides a special parameter to align the chain order.

\section{Discussion}
In this work, we used DCI to solve two problems in the evaluation of protein complex predicted model:
\begin{enumerate}
\item When evaluating a normal multimer for unspecified receptor and ligand, the evaluation of DockQ will become unstable, while DCI score can give an overall and accurate evaluation.

\item When the number of chains increases, batch interface errors are more likely to occur. In this case, TM-score and lDDT used by CASP do not well reflect the overall structure deviation caused by interface prediction errors, where DCI does a significantly better job.
\end{enumerate}
In terms of the DCI algorithm, one issue that needs to be considered is, while DCI reflects batch interface errors, it is likely to be affected by the number of chains and the length of the structure, that is, more or longer chains will magnify the judgment of the algorithm on batch interface errors, resulting in a very low score. Therefore removing these effects from the algorithm is the main goal in the next step.

It is also worth noting that we have given cutoffs for CAPRI classification using DCI, but it is clear that when the number of chains is large, the classification cutoffs are only references, and what is more important is the value of DCI score to evaluate the structural similarity between complexes.

In terms of the application of DCI score, we consider that DCI score is well suited as a reference label or loss function for protein complex model quality assessment (QA) problems. In addition, it will also play a good role in the task of using deep learning models to predict the distance matrix between strands and contact residue pairs.

\section{Conclusion}
In this work we proposed a new assessment metric, DCI for evaluating protein complex predicted model quality. DCI score is a continuous value within 0 to 1, which only depends on the distance maps and contact-interface map of structures. Instead of dividing a complex into two parts (ligand and receptor) like DockQ, DCI can evaluate a structure as a whole. And DCI can better evaluate batch interaface errors in multi-chain case. This accurate and stable evaluation assessment can better guide the prediction task of protein complex, and for protein engineering such as drug design, it also has a positive influence.



\begin{thebibliography}{17}
\providecommand{\natexlab}[1]{#1}
\providecommand{\url}[1]{\texttt{#1}}
\expandafter\ifx\csname urlstyle\endcsname\relax
  \providecommand{\doi}[1]{doi: #1}\else
  \providecommand{\doi}{doi: \begingroup \urlstyle{rm}\Url}\fi

\bibitem[Kabsch(1976)]{1976A}
W.~Kabsch.
\newblock A solution for the best rotation to relate two sets of vectors.
\newblock \emph{Acta Crystallographica Section A}, 32, 1976.

\bibitem[Kabsch(1978)]{Kabsch1978ADO}
Wolfgang Kabsch.
\newblock A discussion of the solution for the best rotation to relate two sets of vectors.
\newblock \emph{Acta Crystallographica Section A}, 34:\penalty0 827--828, 1978.

\bibitem[Zhang and Skolnick(2007)]{TMscore}
Yang Zhang and Jeffrey Skolnick.
\newblock Scoring function for automated assessment of protein structure template quality.
\newblock \emph{Proteins}, 57:\penalty0 702--10, 09 2007.
\newblock \doi{10.1002/prot.20264}.

\bibitem[Zemla(2003)]{PMID:12824330}
Adam Zemla.
\newblock Lga: A method for finding 3d similarities in protein structures.
\newblock \emph{Nucleic acids research}, 31\penalty0 (13):\penalty0 3370—3374, July 2003.
\newblock ISSN 0305-1048.
\newblock \doi{10.1093/nar/gkg571}.
\newblock URL \url{https://europepmc.org/articles/PMC168977}.

\bibitem[Zemla et~al.(2001)Zemla, Venclovas, Moult, and Fidelis]{CASP4}
Adam Zemla, Česlovas Venclovas, John Moult, and Krzysztof Fidelis.
\newblock Processing and evaluation of predictions in casp4.
\newblock \emph{Proteins: Structure, Function, and Bioinformatics}, 45:\penalty0 13 -- 21, 02 2001.
\newblock \doi{10.1002/prot.10052}.

\bibitem[Mariani et~al.(2013)Mariani, Biasini, Barbato, and Schwede]{10.1093/bioinformatics/btt473}
Valerio Mariani, Marco Biasini, Alessandro Barbato, and Torsten Schwede.
\newblock {lDDT: a local superposition-free score for comparing protein structures and models using distance difference tests}.
\newblock \emph{Bioinformatics}, 29\penalty0 (21):\penalty0 2722--2728, 08 2013.
\newblock ISSN 1367-4803.
\newblock \doi{10.1093/bioinformatics/btt473}.
\newblock URL \url{https://doi.org/10.1093/bioinformatics/btt473}.

\bibitem[Gao and Skolnick(2010)]{IS}
Mu~Gao and Jeffrey Skolnick.
\newblock ialign: A method for the structural comparison of protein-protein interfaces.
\newblock \emph{Bioinformatics (Oxford, England)}, 26:\penalty0 2259--65, 09 2010.
\newblock \doi{10.1093/bioinformatics/btq404}.

\bibitem[Jumper et~al.(2021)Jumper, Evans, Pritzel, Green, Figurnov, Ronneberger, Tunyasuvunakool, Bates, Žídek, Potapenko, Bridgland, Meyer, Kohl, Ballard, Cowie, Romera-Paredes, Nikolov, Jain, Adler, and Hassabis]{AF2}
John Jumper, Richard Evans, Alexander Pritzel, Tim Green, Michael Figurnov, Olaf Ronneberger, Kathryn Tunyasuvunakool, Russ Bates, Augustin Žídek, Anna Potapenko, Alex Bridgland, Clemens Meyer, Simon Kohl, Andrew Ballard, Andrew Cowie, Bernardino Romera-Paredes, Stanislav Nikolov, Rishub Jain, Jonas Adler, and Demis Hassabis.
\newblock Highly accurate protein structure prediction with alphafold.
\newblock \emph{Nature}, 596:\penalty0 1--11, 08 2021.
\newblock \doi{10.1038/s41586-021-03819-2}.

\bibitem[Evans et~al.(2021)Evans, O{\textquoteright}Neill, Pritzel, Antropova, Senior, Green, {\v Z}{\'\i}dek, Bates, Blackwell, Yim, Ronneberger, Bodenstein, Zielinski, Bridgland, Potapenko, Cowie, Tunyasuvunakool, Jain, Clancy, Kohli, Jumper, and Hassabis]{AF2-multimer}
Richard Evans, Michael O{\textquoteright}Neill, Alexander Pritzel, Natasha Antropova, Andrew Senior, Tim Green, Augustin {\v Z}{\'\i}dek, Russ Bates, Sam Blackwell, Jason Yim, Olaf Ronneberger, Sebastian Bodenstein, Michal Zielinski, Alex Bridgland, Anna Potapenko, Andrew Cowie, Kathryn Tunyasuvunakool, Rishub Jain, Ellen Clancy, Pushmeet Kohli, John Jumper, and Demis Hassabis.
\newblock Protein complex prediction with alphafold-multimer.
\newblock \emph{bioRxiv}, 2021.
\newblock \doi{10.1101/2021.10.04.463034}.
\newblock URL \url{https://www.biorxiv.org/content/early/2021/10/04/2021.10.04.463034}.

\bibitem[Kryshtafovych et~al.(2021)Kryshtafovych, Schwede, Topf, Fidelis, and Moult]{CASP14_1}
Andriy Kryshtafovych, Torsten Schwede, Maya Topf, Krzysztof Fidelis, and John Moult.
\newblock Critical assessment of methods of protein structure prediction(casp)–round xiv.
\newblock \emph{Proteins: Structure, Function, and Bioinformatics}, 89, 09 2021.
\newblock \doi{10.1002/prot.26237}.

\bibitem[Ozden et~al.(2021)Ozden, Kryshtafovych, and Karaca]{CASP-CAPRI}
Burcu Ozden, Andriy Kryshtafovych, and Ezgi Karaca.
\newblock Assessment of the casp14 assembly predictions.
\newblock \emph{Proteins: Structure, Function, and Bioinformatics}, 89, 08 2021.
\newblock \doi{10.1002/prot.26199}.

\bibitem[Lensink et~al.(2007)Lensink, Mendez~Giraldez, and Wodak]{CAPRI}
Marc Lensink, Raul Mendez~Giraldez, and Shoshana Wodak.
\newblock Docking and scoring protein complexes: Capri 3rd edition.
\newblock \emph{Proteins}, 69:\penalty0 704--18, 12 2007.
\newblock \doi{10.1002/prot.21804}.

\bibitem[Basu and Wallner(2016)]{DockQ}
Sankar Basu and Björn Wallner.
\newblock Dockq: A quality measure for protein-protein docking models.
\newblock \emph{PLoS ONE}, 11:\penalty0 e0161879, 08 2016.
\newblock \doi{10.1371/journal.pone.0161879}.

\bibitem[Hopf et~al.(2017)Hopf, Ingraham, Poelwijk, Schärfe, Springer, Sander, and Marks]{EVcoupling}
Thomas Hopf, John Ingraham, Frank Poelwijk, Charlotta Schärfe, Michael Springer, Chris Sander, and Debora Marks.
\newblock Mutation effects predicted from sequence co-variation.
\newblock \emph{Nature Biotechnology}, 35, 01 2017.
\newblock \doi{10.1038/nbt.3769}.

\bibitem[Yan and Huang(2021)]{10.1093/bib/bbab038}
Yumeng Yan and Sheng-You Huang.
\newblock {Accurate prediction of inter-protein residue-residue contacts for homo-oligomeric protein complexes}.
\newblock \emph{Briefings in Bioinformatics}, 22\penalty0 (5), 03 2021.
\newblock ISSN 1477-4054.
\newblock \doi{10.1093/bib/bbab038}.
\newblock URL \url{https://doi.org/10.1093/bib/bbab038}.
\newblock bbab038.

\bibitem[Frazier(2018)]{Frazier2018ATO}
P.~Frazier.
\newblock A tutorial on bayesian optimization.
\newblock \emph{ArXiv}, abs/1807.02811, 2018.

\bibitem[Callaway(2022)]{casp-af2}
Ewen Callaway.
\newblock After alphafold: protein-folding contest seeks next big breakthrough.
\newblock \emph{Nature}, 613, 12 2022.
\newblock \doi{10.1038/d41586-022-04438-1}.

\end{thebibliography}





\end{document}